\documentclass[allclo,superscriptaddress,eqsecnum,amsfonts,showpacs]{revtex4}
\usepackage{epsfig}

\newcommand{\be}{\begin{equation}}
\newcommand{\ee}{\end{equation}}
\newcommand{\bea}{\begin{eqnarray}}
\newcommand{\eea}{\end{eqnarray}}

\newcommand{\ep}{i\varepsilon}
\newcommand{\nn}{\nonumber}

\begin{document}

\preprint{ \parbox{1.5in}{\leftline{hep-th/??????}}}

\title{On the decoupling solution  for pinch technique gluon propagator}

\author{Vladimir ~\v{S}auli}
\affiliation{CFTP and Dept. of Phys.,
IST, Av. Rovisco Pais, 1049-001 Lisbon,
Portugal, DTP INP Rez near Prague, CAS. }

\begin{abstract}

Within a simple Ansatz for renormalized gluon propagator and using  gauge invariant pinch-technique for Schwinger-Dyson equation,  the limits on the effective gluon mass is derived. We calculated scheme invariant running coupling, which in order to be  well defined, gives the lower limit on the gluon mass. We conclude mass should be larger as $m>0.4\Lambda$ in order to avoid
Landau ghost. The upper limit is estimated from assumed quark mass generation which requires gauge coupling must be  large enough to trigger chiral symmetry breaking. It allows only small range of  $m$, which lead to a reasonably large infrared coupling. Already for $m\simeq \Lambda$ we get no chiral symmetry breaking at all.  
Further, we observe that sometimes assumed or postulated Khallen-Lehmann representation for running coupling is not achieved for any value of $m$. 
\end{abstract}

\maketitle


\section{Introduction}

Infrared behavior of Greens functions (GFs) of Yang-Mills theories has been intensively studied in the last decade. Nonperturbative information about dynamical symmetry breaking and confinement (e.g  free non-propagation of confined degrees of freedom) could be encoded in the low $q^2$ behavior of GFs.  Mostly gauge-variant GFs have been obtained by solving Schwinger-Dyson equations (SDEs) and simulated on the lattice as well. In the case of pure gluodynamics, the   methods based on  truncated  set of SDEs offer recently two type of solutions. As these solutions are non-unique and both type slightly modeled truncation dependent they should be tested consequently  when use to calculate  gauge invariant observable. The so called {\it scaling} solutions have power law momentum behaviour with well defined exponent in the infrared \cite{LESM2002,ZW2002,FIPA2007} and lead to infrared vanishing gluon propagator and correspondingly infrared enhanced ghost propagator. Mesons and baryons can be described by SDEs themselves, actually Bethe-Salpeter and relativistic Fadeev equation are simply the parts of SDEs. Successful matching between gauge invariant hadronic observables and those  gauge variant solutions has not been achieved until now.

Beside of this, there exist the so-called {\it decoupling} solutions
as proposed and showed for instance in \cite{BOU2007,BLYMPQ2008,AGBIPA2008}. Such decoupling solutions typically result in Pinch Technique (PT) SDEs framework \cite{COR1982,COR1986,COR1989,BIPA2002,BIPA2004,BIPA2008} called there massive solution since the massless pole of gluon propagator  disappear and gluon propagator is finite (but nonzero) in the infrared (for the topical review see \cite{rewiev2009}).
Recall, PT rearranges the original gauge scheme dependent GFs in a unique way such that unphysical degrees of freedom are eliminated.  In the mean time, as the lattice calculations \cite{lat1,lat2,lat3} in conventional Landau gauge start to support the decoupling solution reaching recently quite infrared momenta down to 75 MeV \cite{lat4}, it attracts new attention again \cite{PENIN,cornwall2009}.

Principal advantage of the PT is its scheme and gauge invariance.
It has been proved to  all orders of perturbation theory that 
the  PT GFs   satisfy Ward identities 
and that extracted effective charges are process independent. Furthermore, GFs do not depend on a gauge fixing parameters. From this automatically follows that  when hadron property are calculated within the QCD PT GFs, an unphysical degrees freedom are eliminated form the beginning. 
 However, as always necessary, the approximation must be made to truncate the infinite tower of the SDEs series. In the original paper the gauge technique was used to construct gluon vertex. To that point Kallen-Lehmann representation (KLR) was assumed for the gluon propagator.  
In contradiction, in other studies it is  suggested  that instead of validity, this is just the  absence of KLR which can ensure confinement \cite{alkofer}. Pure Yang-Mills part of  QCD could be responsible for confinement, therefore the gauge technique construction based intrinsically on the KLR  may be actually a  weak point of the PT.  To that point we recall that in  the later stage \cite{COR1986,COR1989} the gauge technique  was omitted and the gluon  vertex were constructed independently on KLR \cite{COR1989}. In this paper, we do not consider KLR as a reasonable criterion, instead of as a meaningful criterion for the gluon propagator behavior we consider chiral symmetry breaking which must be triggered by this propagator in usual sense. Clearly, the appearance of  chiral symmetry breaking is one of the main "musts" of QCD.

In the next section we briefly rederive the SDE with parametrization of the solution as made in 
\cite{cornwall2009} and basic ingredients are reviewed in this section for completeness. In the third Section we find the discuss this solution and find the restrictions  which must the resulting propagator obey. 

\section{PT SDE with gauge invariant vertex}

In the paper \cite{cornwall2009} the PT  propagator based on the WTI improved vertex \cite{COR1989}  was considered.
Making a simple parametrization  of the solution then the SDE has been  solved analytically.
We adopt here the form of the solution proposed in  \cite{cornwall2009} and obtain the running coupling for all $q^2$. 
  
The product of the gauge coupling  $g^2$ and PT  gluon propagator $\hat{d}$
defines renormalization invariants. It is certainly allowed to rewrite this product by a new one, where one function can represent invariant running charge while the second, say the function $H$ stays for the rest, let us assume  the function $H$ shows up a massive pole instead massless one  
. Using the same convention as in \cite{cornwall2009} we can write:
\be \label{prd}
g^2\hat{d}(q^2)=\bar{g}^2(q^2)\hat{H}(q^2)
\ee
Clearly the functions on rhs. of Eq. (\ref{prd}) are obviously not uniquely define 
if one does not say more. This problem is simply avoided if one assume the form of $H$ explicitly
since then only one $\bar{g}^2(q^2)$ function needs to be identified.

The simplest parametrization we can use is the following  hard mass approximation
\be 
g^2\hat{d}(q^2)=\bar{g}^2(q^2)\frac{1}{q^2-m^2+\ep}.
\ee

In some approaches \cite{SHISOL2007} it is assumed that this is the running coupling
which can satisfy Khallen-Lehmann representation (KLR).
In paper \cite{cornwall2009} it is conjectured that the both functions $\bar{g}^2$ and $\hat{H}$  
satisfy KLR, so the  integral representation for the product has the same analyticity domain but 
the absorptive part is not positive semidefinite. It should be negative in places in accordance with the one loop ultraviolet  asymptotic  $g^2\hat{d}(q^2)\simeq 1/[q^2ln(q^2)]$.

Remind here, KLR for the running charge should  read 
\be \label{KLR}
\bar{g}^2_{KLR}(q^2)=\frac{1}{\pi}\int d\omega \frac{\Im \bar{g}^2_{KLR}(\omega)}{q^2-\omega+\ep}\, . 
\ee
Such running  function  is homomorphic  in whole complex plane up to the real positive semi-axis of $q^2$ where the branch points are located.

The PT SDE is represented by a non-linear integral equation derived in \cite{COR1986} and solved first time in \cite{cornwall2009}, it reads
\be \label{ptsde}
\left[\bar{g}^2\hat{d}(q^2)\right]^{-1}=q^2bZ-\frac{ib}{\pi^2}\int d^{4k}\hat{H}(k)\hat{H}(k+q)\left[q^2+\frac{m^2}{11}\right]+C \,\, , 
\ee
where $C$ is momentum independent constant and where as the hard mass approximation has been  employed. The one loop  beta function coefficient  is   
\be
b=\frac{11N_c-2N_f}{48\pi^2}.
\ee

After the renormalization, which was made by on shell-subtraction (note, renormalization is  not multiplicative here, for details see \cite{cornwall2009}) we get for Eq. (\ref{ptsde})
\be \label{rptsde}
\left[\bar{g}^2\hat{d}
(q^2)\right]^{-1}=b\left[J(q)(q^2+\frac{m^2}{11})-J(m)\frac{12m^2}{11}\right]\, ,
\ee
where the function $J$ is renormalized in  accordance with the correct one loop ultraviolet asymptotic and it reads
\be
J(q)=-\int_{4m^2}^{\infty}d \omega\frac{q^2}{\omega}\frac{\rho(\omega;m)}{q^2-\omega+\ep}+2+2\ln{(m/\Lambda)}
\ee
where $\Lambda$ is usual QCD scale   valued few hundred $MeV$ for $N_f=2$
and $\rho(\omega;m)=\sqrt{1-\frac{4m^2}{\omega}}$
. 
The integral is a textbook scalar 1-loop integral and can be easily evaluated as following:
\bea
J(q)&=&\rho\ln{\left|\frac{1+\rho}{1-\rho}\right|}+2\ln(m/\Lambda)  
\nn \\
&-&i\pi\rho\theta(q^2-4m^2)\,\; 
\, \, \,\,\,\,\,\,\mbox{for} \, \,  1-4m^2/q^2>0
\nn \\
J(q)&=&-i2\rho{\mbox{arctg}}\left(\frac{i}{\rho}\right)+2\ln(m/\Lambda)\,; \, \,\, \, \,\mbox{for} \, \, 0<q^2<4m^2  \, .
\eea
where $\rho=\sqrt{1-\frac{4m^2}{q^2}}$.

\section{Physically admissible solutions}

Inverting the SDE (\ref{rptsde}),  the pinch technique gauge invariant running charge can be straightforwardly evaluated:
\be
b4\pi\alpha_{massive}(q^2)=b\bar{g}^2(q^2)=\frac{q^2-m^2}{J(q)(q^2+\frac{m^2}{11})-I(m)\frac{12m^2}{11}}
\ee

The imaginary part of $J$ for $q^2>4m^2$ is given by very simple phase space factor $\pi\rho=\pi(1-4m^2/q^2)$, thus the imaginary and the real  parts of the running coupling read
\bea \label{icka}
Im \, b\bar{g}^2&=&(1-\gamma)\frac{\pi\rho(q^2)}{(Re J(q)-\gamma J(m))^2+(\pi\rho(q^2))^2}\,  \, ;
\nn \\
Re \, b\bar{g}^2&=&(1-\gamma)\frac{Re J(q)-\gamma J(m)}{(Re J(q)-\gamma J(m))^2+(\pi\rho(q^2))^2} \, ,
\eea
where we have used the short notation $\gamma=\frac{12m^2}{11q^2+m^2}$. For $q^2<4m^2$ the imaginary part vanishes.

As our approximated PT gluon propagator has a single  real pole, it does not respect confinement issues. Nevertheless we expect that this is a good approximation of the true full (exact) PT solution for the gluon propagator which could  posses a large enhancement at vicinity of $q^2\simeq m^2$ instead of the real pole we use. The singularity of exact PT propagator need are very likely situated away of the real axis and need not be a simple pole but a branch point(s). Furthermore,  we argue that for any $m$ the resulting $\alpha$ does not satisfy KLR. Actually although high $s=q^2$ behaviour of absorptive part exactly corresponds with known analyticized one loop QCD coupling \cite{SHIRKOV,MISO1997,MISO1998}. i.e. for 
$s>>m^2,\Lambda^2$ we can get
\be
Im \alpha(s) \rightarrow \frac{4\pi/b}{\pi^2+\ln^2(s/\Lambda^2)},
\ee
but this the appearance of nontrivial threshold which crucially changes  real part of the running coupling behaviour when $m$ is nonzero. Before showing this explicitly we discuss general features of the solution.

Since we have basically reproduced  Cornwall's solution \cite{cornwall2009} for a general $m$ ,  we  will briefly summarize results for completeness here. By construction, the correct one loop perturbation theory  behaviour is reproduced for any $m$, i.e. $\alpha\simeq 1/ln(q^2/\Lambda^2)$ for $q^2>>\Lambda,m$. There exists  certain critical mass, say $m_c^{I}$, bellow which the coupling $\alpha$ is more singular  as it posses  nonsimple pole,  this critical mass is approximately given by the ratio $m_c^{I}/\Lambda=1.2$ here. Decreasing the mass parameter  $m$ we can find second critical point, say $m_c^{II}$ ,where this pole crosses Minkowski light cone $q^2=0$ and becomes well known unphysical spacelike Landau pole. This  Landau ghost is really unacceptable and it gives severe boundary on the gluon mass, in our case we get approximately  $m_c^{II}=0.4\Lambda$.     Examples of running coupling  are shown in  Fig. 1 for various $m$. Singularities move to the left as $m$ decrease. For large $m$ enough,
$m>1.2\Lambda$, the only  standard two particle branch point singularity located at $4m^2$ remains and the pole has gone. The  singularity can appear only under the threshold, wherein the zero of $\alpha^{-1}$ is not protected by a nonzero absorptive part. As an exotic solution, there also exists a "double pole" (note, it is not a simple double pole) solution for a specific $m$, i.e. the pole in $H$ is enhanced by non simple pole of $\alpha$ at the same point $q=m$. 

\begin{figure}
\centerline{\epsfig{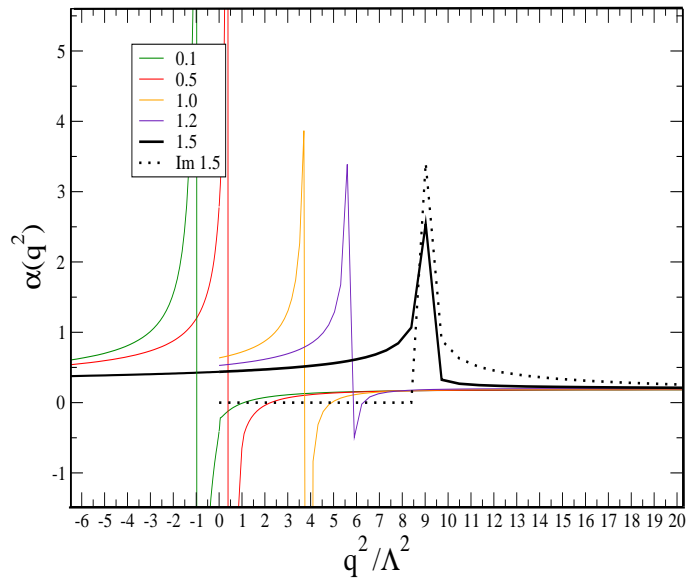}}
\caption[caption]{Pinch technique coupling $\alpha$ for a various ratio $m/\Lambda$. For better identification $Im $ part is displayed for regular solution with $m/\Lambda=1.2$ only.  } 
\end{figure}

The gluon propagator vanishes faster then $1/q^2$ , thus it cannot satisfy KLR. It was suggested in the paper \cite{cornwall2009} thatthe PT gluon propagator can the product of the coupling and newly introduced  function $H$ both of them satisfying their own KLR. Due to this reasoning the author excluded such solutions which lead to the singular coupling in the timelike region as well. As a matter of the fact we argue that the pinch technique running coupling does not satisfy KLR for any $m$, even for those values where $\alpha$ is regular. We check wether the coupling   $\alpha^{KLR}(q^2)=\bar{g}^2_{KLR}(q^2)/(4\pi)$ is analytical in usual sense, since it must  satisfy KLR (\ref{KLR}). It has been done by substitution of  the absorptive part of $\alpha^{PT}$ into rhs. of (\ref{KLR}) , subsequently the  obtained  $\alpha^{KLR}(q^2)$ is compared with the  real part of  $\alpha^{PT}$  already known from   the solution of PT SDE. For easiest inspection we choose spacelike $q^2$ where the dispersion integral is regular and $\alpha^{KLR}(q^2)$ can be evaluated with arbitrary accuracy. The comparison of analyticized coupling with $\alpha^{PT}$ is shown in  Fig.2. For any $m$ the coupling $\alpha^{PT}$ never agrees with "analyticized" coupling defined by  (\ref{KLR}). For instance for $m=1.5$, where  the best optimized approximation is roughly achieved, one gets 30 $\%$ underestimation in the infrared, while we get complete disagreement for large $q^2$.  Clearly, the dispersion relation, which would be very usefull in other practical calculations, is not followed here. We argue that there is no any known physical reason to expect KLR for the running coupling.
 Instead of, as $\alpha$  is the unique form factor of the pinch gluon propagator its absence can  regarded as consequence of the confinement.

\begin{figure}
\centerline{\epsfig{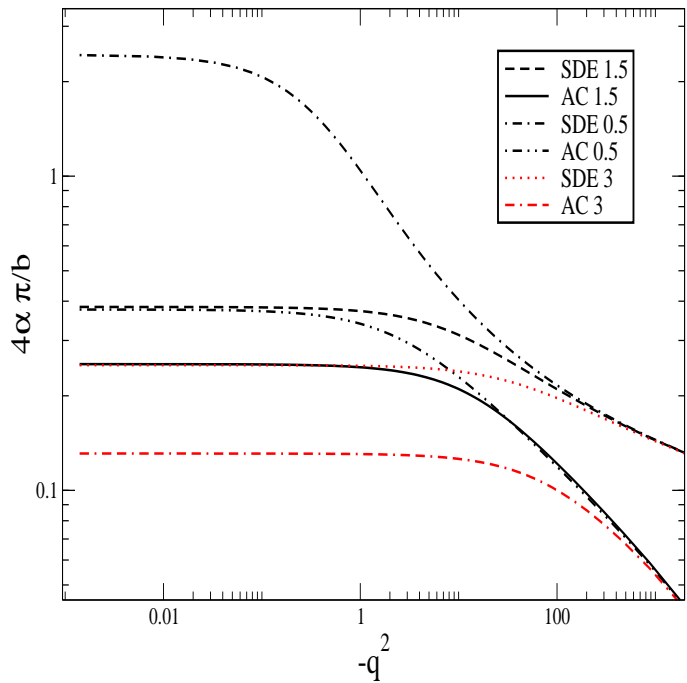}}
\hspace{1cm}
\caption[caption]{$b\bar{g^2}$ plotted for spacelike fourmomenta for various ratio of $m$. It is compared with analyticized coupling (AC) as described in the text.} 
\end{figure}

 Let us mention here that the absence of KLR for propagator  is perhaps common phenomenon of a strong coupling theory and it is quite independent on a model details.
Actually, the absence of KLR has been already observed in strong coupling QED, scalar toy models \cite{SAULI} and in the case of quark propagator in  Landau gauge QCD \cite{SAUBI}.
Roughly say, to get  the dispersion relation for selfenergy, i.e. for the inverse of the propagator, and simultaneously expect the spectral representation for the propagator itself is too strong assumption in the strong coupling theory like QCD.

  As we discussed,  mass $m$ is severely constrained  from bellow by requirement of absence of the spacelike Landau pole. For an indication of what $m$ should be we do not choose the criterion of KLR, which appear as a quite obscured requirement in confining theory, but we require that pinch technique running coupling must be large enough in order to trigger correct chiral symmetry breaking in QCD.  In the real QCD the dynamical chiral symmetry breaking is phenomenon responsible for the most of nucleon mass (i.e. for the   u,d quarks dynamical mass generation) while it it simultaneously explains the   lightness of the pions.  To describe all these observable in selfcontained way, the chiral symmetry breaking must be correctly incorporated into the formalism. Such requirement very naturally gives upper boundary on the pinch technique gluon mass since for  gluon heavy enough,  the pinch technique running coupling is too weak and it does not trigger chiral symmetry breaking. 

In principle, employing the formalism of PT SDEs for gluons and quarks simultaneously solved with  the Bethe-Salpeter equations  one should be able  to fit the mass in  the PT gluon propagator from meson spectra. Unhappily, the recent calculations are still far  from this stage even in the more conventional gauge fixed schemes and the form of propagators entering the calculation is still dubious. First, we will describe arising  obstacles of such a treatment and we suggest simplified way to make an reliable estimate of $m$, which is solely based on the solution of quark gap equation in the ladder approximation. 

In lattice QCD,  a static quark-antiquark potential can be computed with the Wilson loop technique. This gives us confining linear potential $V_L$ between infinitely heavy quarks. For a correct description of excited mesons this  could be principally involved covariantly in the quark-antiquark kernel of BSE. From the other side the  various hadronic observables were calculated in the framework of Schwinger-Dyson equations during last two decades, it includes the meson spectra and decays \cite{meson1997,meson2007,meson2008} and various form factors \cite{ff2000,ff2008}, more complicated baryonic properties were studied in this framework as well. Most of them use the ladder approximation of the quark SDE (and meson BSE) in Landau gauge, while first steps beyond  the ladder approximation has been considered only quite recently. Recall, the ladder approximation means that "very effective" one  gluon exchange is considered only, while a more gluon exchanges are needed and more topologically more complicated
diagrams must contribute to get linear potential in non-relativistic limit. 
Such higher skeletons quite naturally generates important scalar part of the quark-antiquark potential \cite{Bicudo:2003ji}.  
We conjecture that if the quark-antiquark BSE kernel analogue of $V_L$ is not included then this is the main source of discrepancy when one compare GFs used in meson ladder calculations and the GFs actually obtained from SDEs. It is more then obvious that the effective gluon propagator used in a typical ladder approximated BSE more or less models  unconsidered higher order skeletons. Without reasonable matching of SDEs calculations on gluon propagator and  the one entering kernels  of the meson BSE, the infrared behaviour of gluon propagator is not obvious.

On the other hand, our knowledge of Wilson lattice results combined with the knowledge of typical quark SDE solutions offers  economical way, which we argue is efficient enough in order to estimate the PT gluon propagator in the infrared. For this purpose , let us consider the solutions of the quark SDE when one gluon and one gluon plus infrared enhanced  effective interaction $V_L$ is added. The difference of these two aolution has been  studied in the paper \cite{BIMACACAOL2009} and it gives approximative double enhancement of the quark dynamical mass in deep infrared Euclidean $Q^2$ when $V_L$ is taken into account. Using these  arguments, the main issue is that infrared quark mass $M(0)$ should be already as large as $M(0)\simeq \Lambda=250 MeV$ when one uses the ladder quark gap equation alone but now with the PT running coupling implemented in. The additional unconsidered term $V_L$ could be then responsible for an additional grow of the quark mass in the same order. It automatically gives the limit on  the running coupling , its value must be significantly larger then the critical coupling, bellow which there is no symmetry breaking at all. As the dynamical quark mass function obtained in the ladder approximation is quite universal, we did need to perform a detailed numerical analyzes of the quark gap equation with PT running coupling and  we can  estimate the solution from the infrared value of the coupling, which must be $\alpha(0) \simeq 2.0$ or larger. To get such value we can see that we must use the solution with $m/\Lambda=0.4-0.7$. Since required interval lies between  $m^I_c$ and $m^{II}_c$ we always have the running coupling singularity at the timelike regime. In this way the pinch  technique offers possible  scenario for Infrared Slavery again, albeit with coupling enhancement in the timelike region simply  due to the massiveness of the gluon.

\section{Conclusion}

Using recently obtained pinch technique gluon propagator   \cite{cornwall2009}  the  limits on the effective gluon mass have been reconsidered.  It is confirmed that in order  to avoid unphysically singular running coupling, the gluon mass must be bounded from bellow. We argue that the requirement of KLR for the running coupling is not a good guide for this purpose and we assume that running coupling can be enhanced or even singular  in the timelike region of the momenta.
It is suggested that the upper boundary on the gluon mass $m$ stems from  chiral symmetry breaking when quarks are considered as well. As the infrared enhancement of the interaction is necessary to get correct triggering of symmetry breaking and since the running coupling crucially depends on mass $m$, the upper boundary stems from the minimal required pinch technique running coupling.
 It gives the acceptable region of the gluon mass $m\simeq 0.4-0.7 \Lambda $, or so. This is in reasonable agreement with the recent lattice results 
and simultaneously it does not contradict the existence of chiral symmetry breaking in QCD.

\end{document}